\documentclass[twocolumn,twoside,slac_two]{revtex4}
\usepackage{graphicx}
\usepackage{fancyhdr}
\begin{document}

\title{Investigation of the hadronic interaction models using WILLI detector}

%

\author{B. Mitrica, I.  M. Brancus, M. Petcu, A. Saftoiu, G. Toma, M. Duma}
\affiliation{National Institute of Physics and Nuclear Engineering – Horia Hulubei, POB MG 6, Ro-76900 Bucharest, Romania}
\author{H. Rebel, A. Haungs}
\affiliation{Karlsruhe Institute of Technology - KIT - Campus North, Institut f¨ur Kernphysik, POB 3640, 76021 Karlsruhe, Germany}
\author{O. Sima}
\affiliation{Department of Physics, University of Bucharest, P.O.B. MG-11, Romania}

\begin{abstract}
The WILLI detector, built in IFIN-HH Bucharest, in collaboration with KIT Karlsruhe, is a rotatable modular detector for measuring charge ratio for cosmic muons with energy $<$ 1 GeV. It is under construction a mini-array for measuring the muon charge ratio in Extensive Air Showers. The EAS simulations have been performed with CORSIKA code. The values of the muon flux, calculated with semi-analytical formula, and simulated with CORSIKA code, based on DPMJET and QGSJET models for the hadronic interactions, are compared with the experimental data determined with WILLI detector. No significant differences between the two models and experimental data are observed. The measurements of the muon charge ratio for different angles-of-incidence, (performed with WILLI detector) shows an asymmetry due to the influence of magnetic field on muons trajectory; the values are in agreement with the simulations based on DPMJET hadronic interaction model. The simulations of muon charge ratio in EAS performed with CORSIKA code based on three hadronic interaction models (QGSJET2, EPOS and SYBILL) show relative small difference between models for H and for the Fe showers; the effect is more pronounced at higher inclination of WILLI detector. The future measurements should indicate which model is suitable.

\end{abstract}

\maketitle

\thispagestyle{fancy}


\section{WILLI detector}
\begin{figure}[h]
\includegraphics[width=40mm]{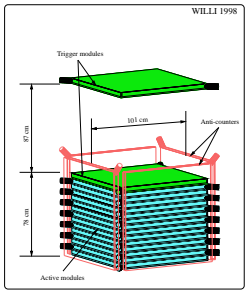}
\includegraphics[width=40mm]{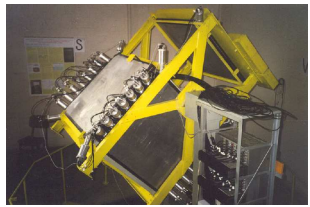}
\caption{Left: The vertical WILLI detector
Right: The rotatable configuration of WILLI} \label{willi}
\end{figure}

The rotatable WILLI detector \cite{1}, shown in Fig.\ref{willi}. is a modular system, each
module being formed by a scintillator layer, 3 cm thickness in Al frame, 1 cm
thickness, operating for all azimuth angles and down to 45$^{0}$ zenith inclination.

The detector permits the measurement of the energy deposited by traversing
muons and additional, for muons stopped in the detector, the moment of decay
up to 50 $\mu$s. The active layers of the detector are plates of scintillators of 1 m$^{2}$
which are read by two photomultipliers mounted on opposite corners of each plate,
using alternatively corners. The main aim of WILLI detector is to determine the
muon charge ratio by measuring the life time of stopped muons in the detector
layers: the stopped positive muons decay with a lifetime of 2.2 $\mu$s, while negative
muons are captured in the atomic orbits, leading to an effectively smaller lifetime
depending on the stopping material. The muon charge ratio is determined from
the measured decay curve of all muons stopped in the detector, by fitting the
measured decay spectrum with the theoretical curve.

\section{The simulations of the cosmic muon flux}

The simulations of the EAS have been performed with CORSIKA program \cite{2},
based on six different models for the description of the high - energy hadronic
interaction: DPMJET II.5 \cite{3}, QGSJET \cite{4}, VENUS \cite{5}, SIBYLL \cite{6} and
three different models for the description of low - energy hadronic interaction:
GHEISHA \cite{7}, UrQMD 1.1 \cite{8} and DPMJET, which includes some extensions
allowing the simulation of hadronic interaction down to 1 GeV; the threshold is
set by default to Elab = 80 GeV/n. The local geomagnetic field is included in
CORSIKA in the approximation of homogeneous field, as described by the International
Geomagnetic Reference Field for the year 2000 \cite{9}. The geomagnetic
cutoff is calculated with a Monte Carlo procedure of the possible particle trajectories
in the back-tracking method, enabling the calculation of a table of allowed
and forbidden trajectories \cite{10}. The particle tracking is based on GEANT 3.21,
\cite{11} starting at 112.83 km, the top of the atmosphere as defined in CORSIKA.

\section{The investigation of the cosmic muon flux}

\begin{figure}[h]
\includegraphics[width=40mm]{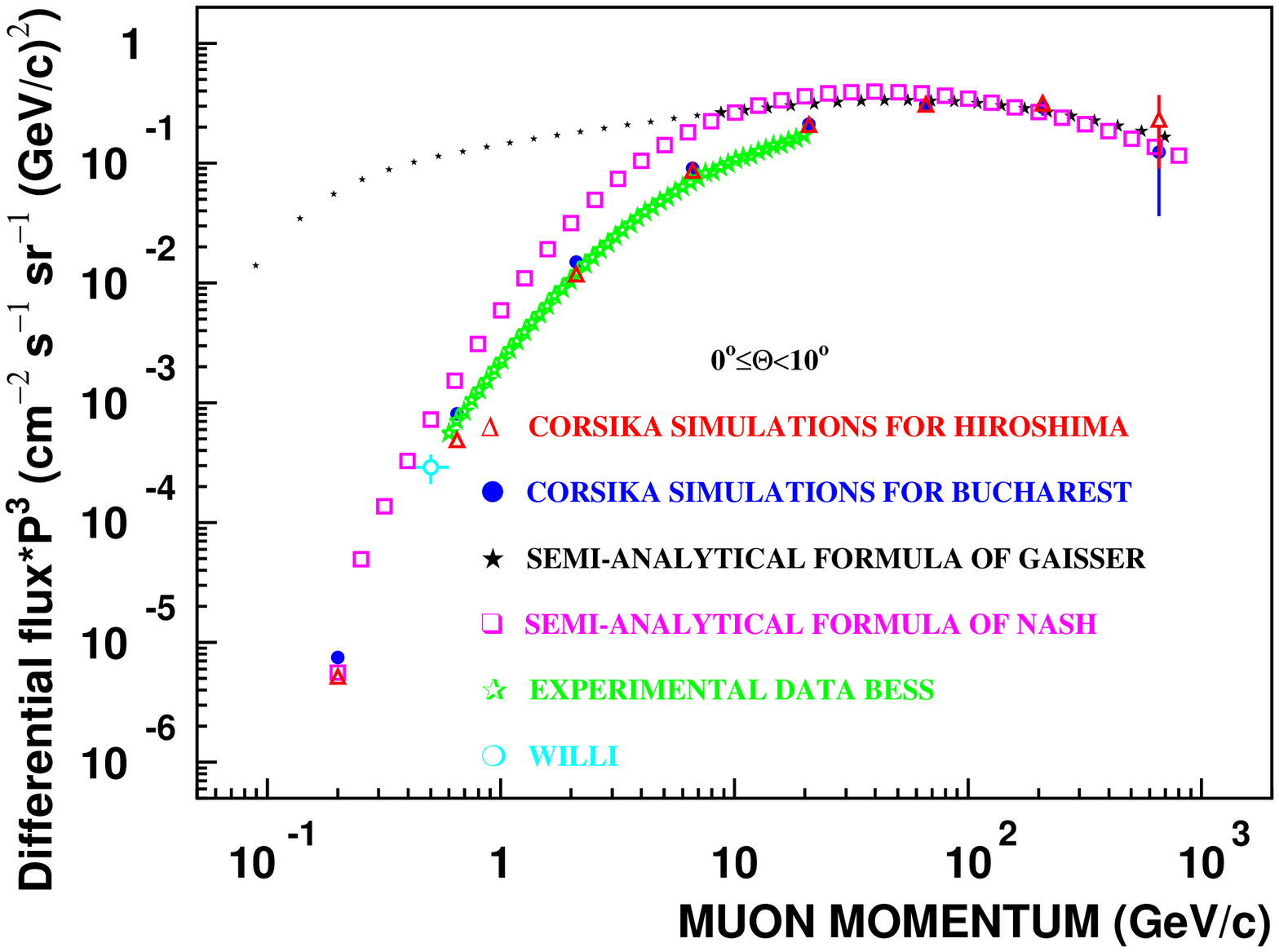}
\includegraphics[width=40mm]{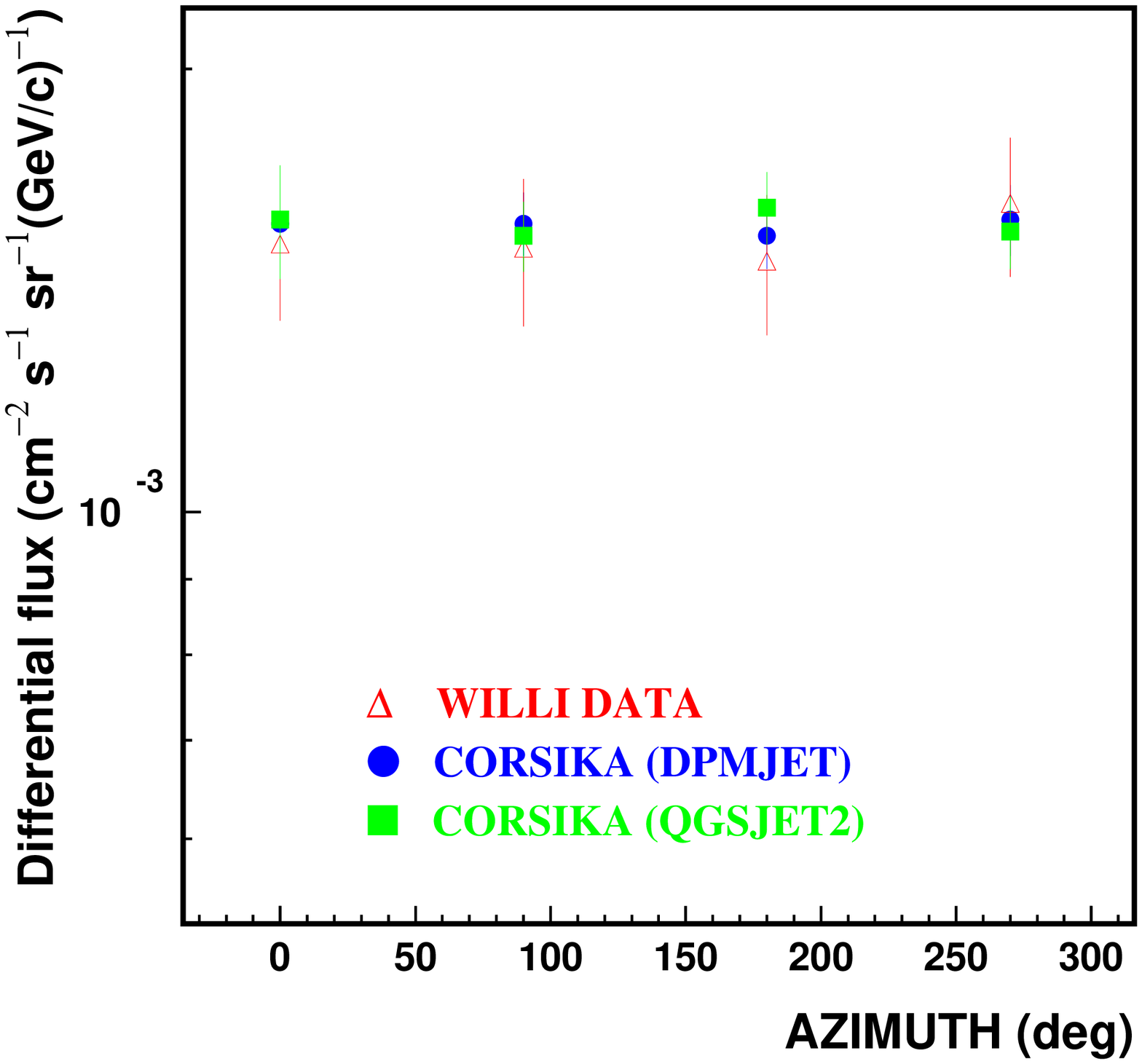}
\caption{The muon flux data compared with Monte Carlo
simulations and semi-analytical formulae for Hirosima and
Bucharest (left); The azimuthal variation of muon flux measured
by WILLI compared with CORSIKA simulations(right)} \label{flux}
\end{figure}

In view of detailing geomagnetic effects, the muon fluxes have been calculated
\cite{12} for two different locations with different magnetic cutoff: Hiroshima (34$^{0}$
N, 132$^{0}$ E) with the geomagnetic cutoff 11.6 GV and Bucharest (44$^{0}$ N, 26$^{0}$ E)
with geomagnetic cutoff of 5.6 GV \cite{10}. Fig.\ref{flux} (left) compares for Bucharest and
Hiroshima the muon flux calculated with CORSIKA with the semi-analytical
formulae of Nash \cite{13} and Geiser \cite{14}.

We see that at low muon energies ( $<$ 1 GeV) the muon flux is significantly
influenced by the geomagnetic field and dependent on the particular observation
site, and from the direction of muon incidence. The muon flux for incident
energies $<$ 1 GeV was compared with CORSIKA simulations using DPMJET
and QGSJET2 models for hadronic interactions, see Fig.\ref{flux} (right).

The simulated data for atmospheric muons obtained with CORSIKA code, using
both DPMJET and QGSJET interaction models, are in good agreement with
measured data of WILLI detector for the whole azimuth range and a mean zenith
angles of 35$^{0}$.


%



\section{The investigation of the charge ratio for the cosmic muons}

\begin{figure}[h]
\includegraphics[width=40mm]{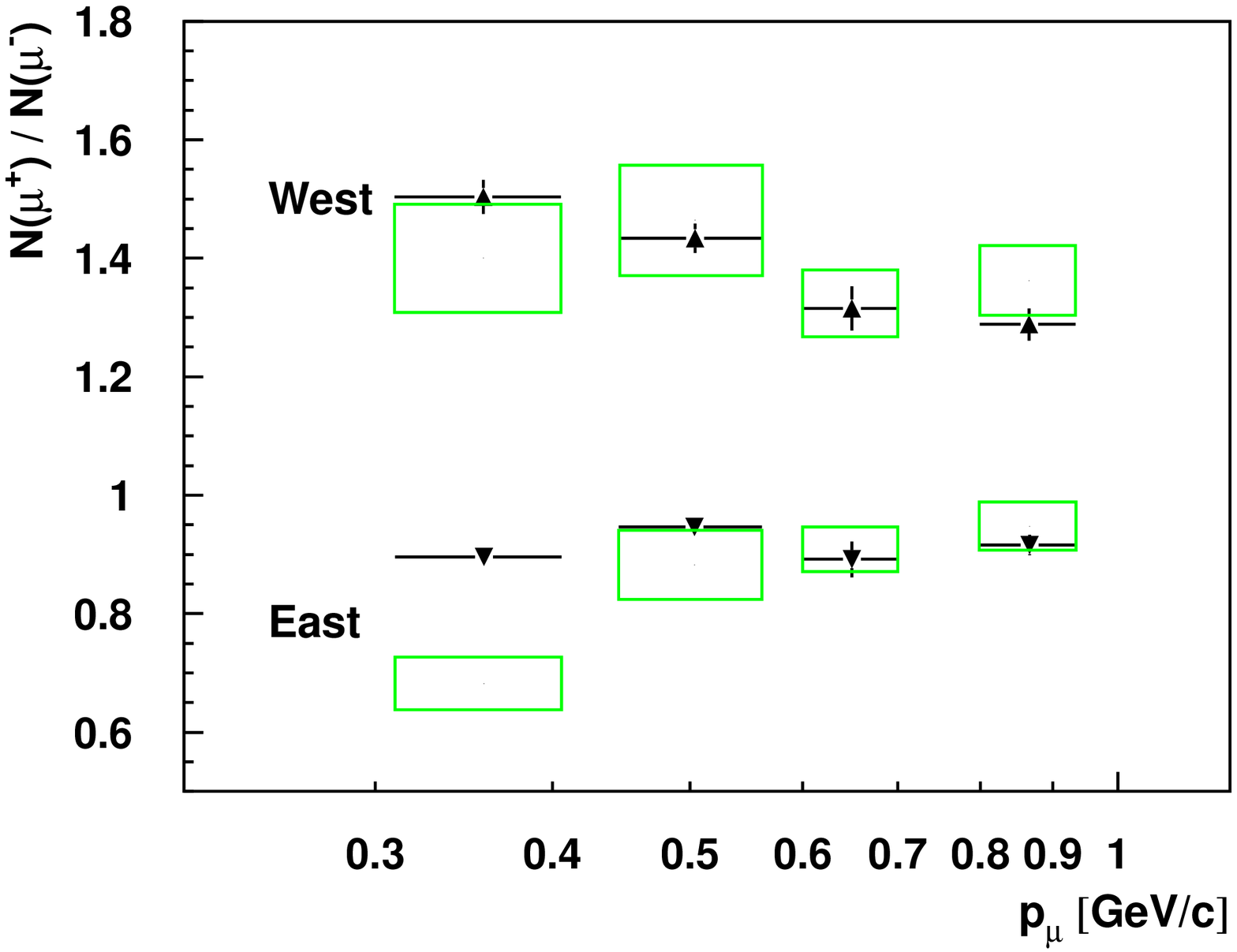}
\includegraphics[width=40mm]{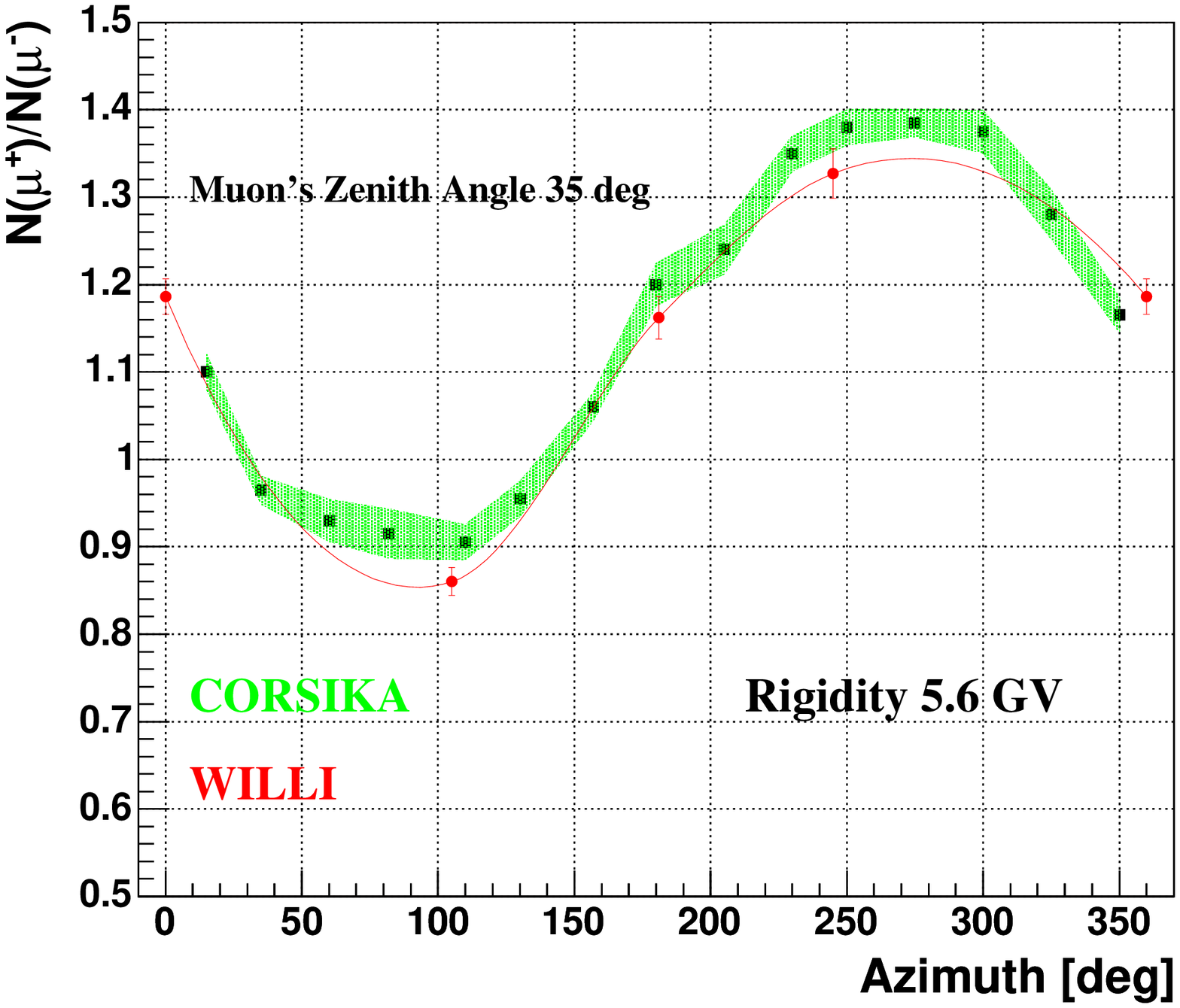}
\caption{Energy dependence of the muon charge ratio, measured
separately for East and West directions, with comparing
experimental data and simulations (left); The azimuthal
variation of the muon charge ratio (right)} \label{charge_ratio}
\end{figure}

The charge ratio R$_{\mu}$($\mu^{+}$/$\mu^{-}$) of the atmospheric muons provides a sensitive
test of the simulation of the fluxes as different charges have different path lengths
from the production level to the detector, and consequently the decay probability
for low-energy muons is modified. This influence leads to the latitude effects of
the flux and to the so-called East West effect \cite{15}, which is more pronounced
at lower muon energies and with larger observation angles, see Fig.\ref{charge_ratio} (left). 

For the investigation of the azimuthal dependence of the  charge ratio of atmospheric muons, a series of measurements \cite{12} has been performed on four azimuth directions of incidence of the atmospheric muons: North, East, South, West, (N, E, S, W) for muons with inclined incidence, mean value at 35$^o$ and mean incident energy  0.5 GeV/c. The results are displayed in Fig.\ref{charge_ratio} (right). 


The good agreement between the measurements and simulations data
indicates that CORSIKA code describes well the azimuth variation i.e. the
East-West effect observed by WILLI. The Okayama group reported less
pronounced azimuth dependence \cite{16} for muon incident energies above
1 GeV. 


\section{The detector WILLI-EAS}

\begin{figure}[h]
\includegraphics[width=40mm]{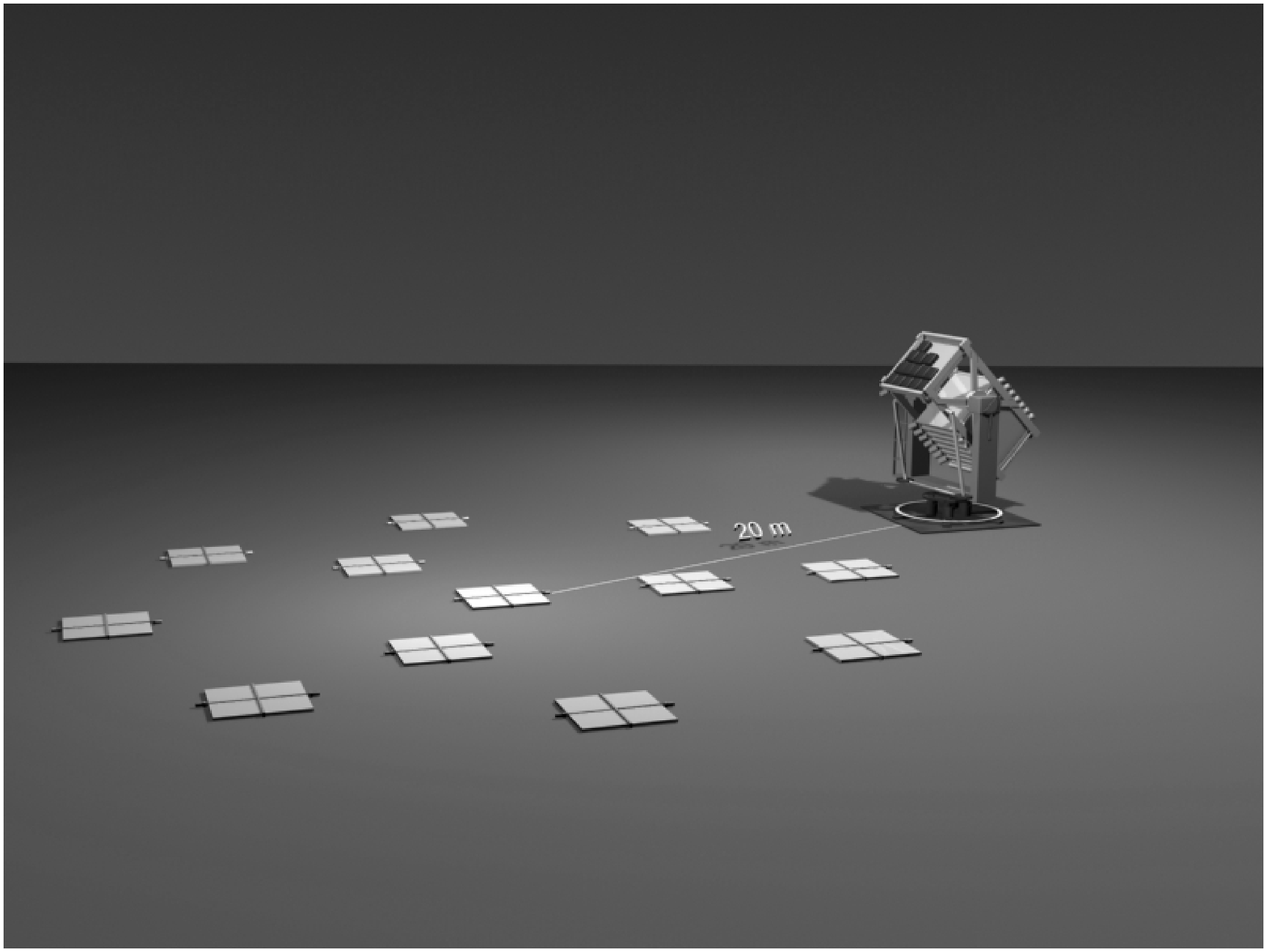}
\includegraphics[width=40mm]{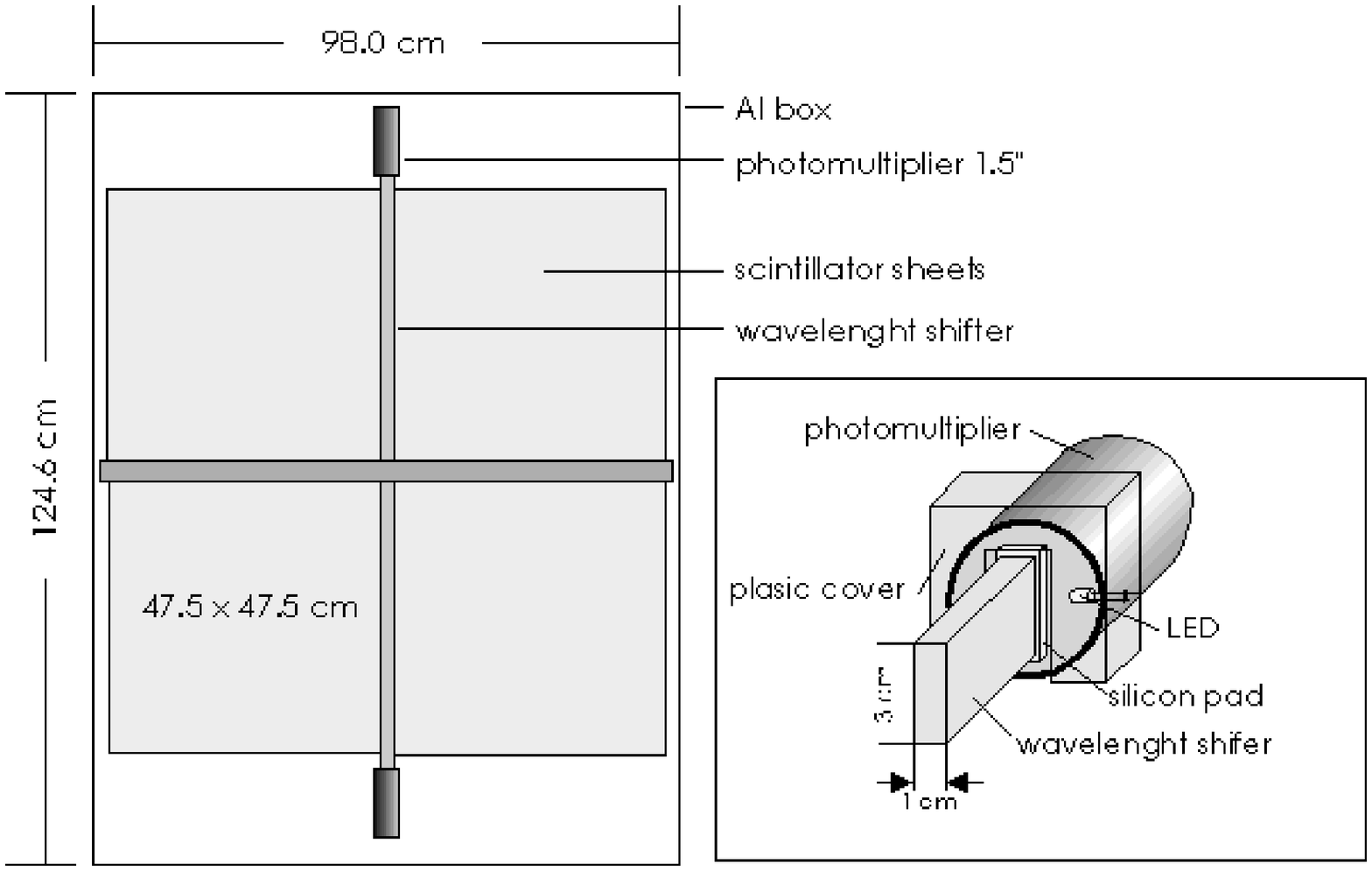}
\caption{WILLI-EAS detector (left); The unit of the mini-array for detecting the shower
events (right)} \label{willi_eas}
\end{figure}

We are building a small scintillator detector array nearby WILLI in order
to trigger the muon detection of WILLI by events of small showers 
and to determine the average charge ratio within
EAS \cite{17}. The mini-array is formed by 12 detector stations, each being
a scintillator plate, 3 cm thickness, divided in 4 parts (0.475 x 0.475 m2 -
see Fig.\ref{willi_eas}).

\begin{figure}[h]
\includegraphics[width=50mm]{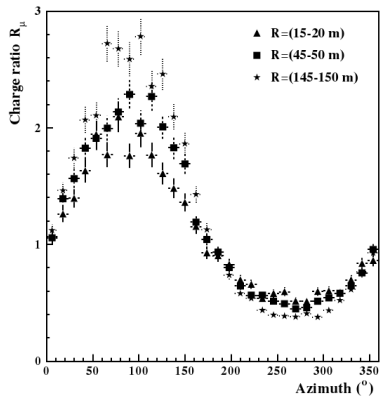}
\caption{Variation of the charge ratio R$_{\mu}$ of the mean muon
density distribution of proton-induced EAS of inclined
showers ($\theta$ = 45$^{0}$) incident from the North with the
primary energy of 10$^{15}$ eV at various distances R from the
shower axis} \label{distance}
\end{figure}

It is expected that like for the East -West effect of atmospheric muons,
the mean charge ratio is affected by the geomagnetic field, especially when
- like it is the case with WILLI - low energy muons are registered. The
simulations of muon charge ratio performed with CORSIKA based on
QGSJET model performed for H, see Fig.\ref{distance}, and Fe showers with primary
energy 10$^{15}$ eV, show azimuthal variation, more pronounced with
increasing distances of WILLI detector, from the shower core \cite{18}.

\begin{figure}[h]
\includegraphics[width=40mm]{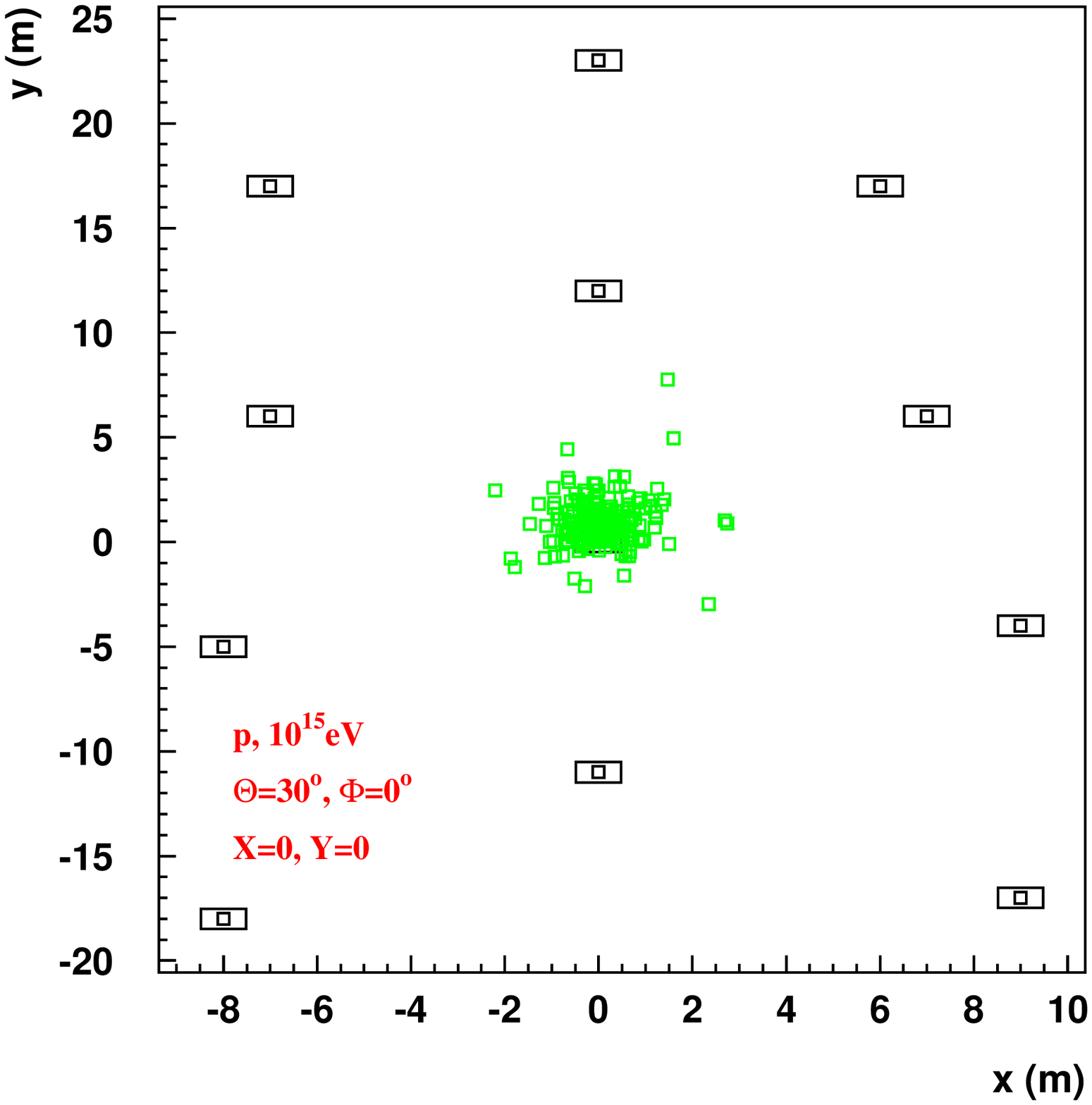}
\includegraphics[width=40mm]{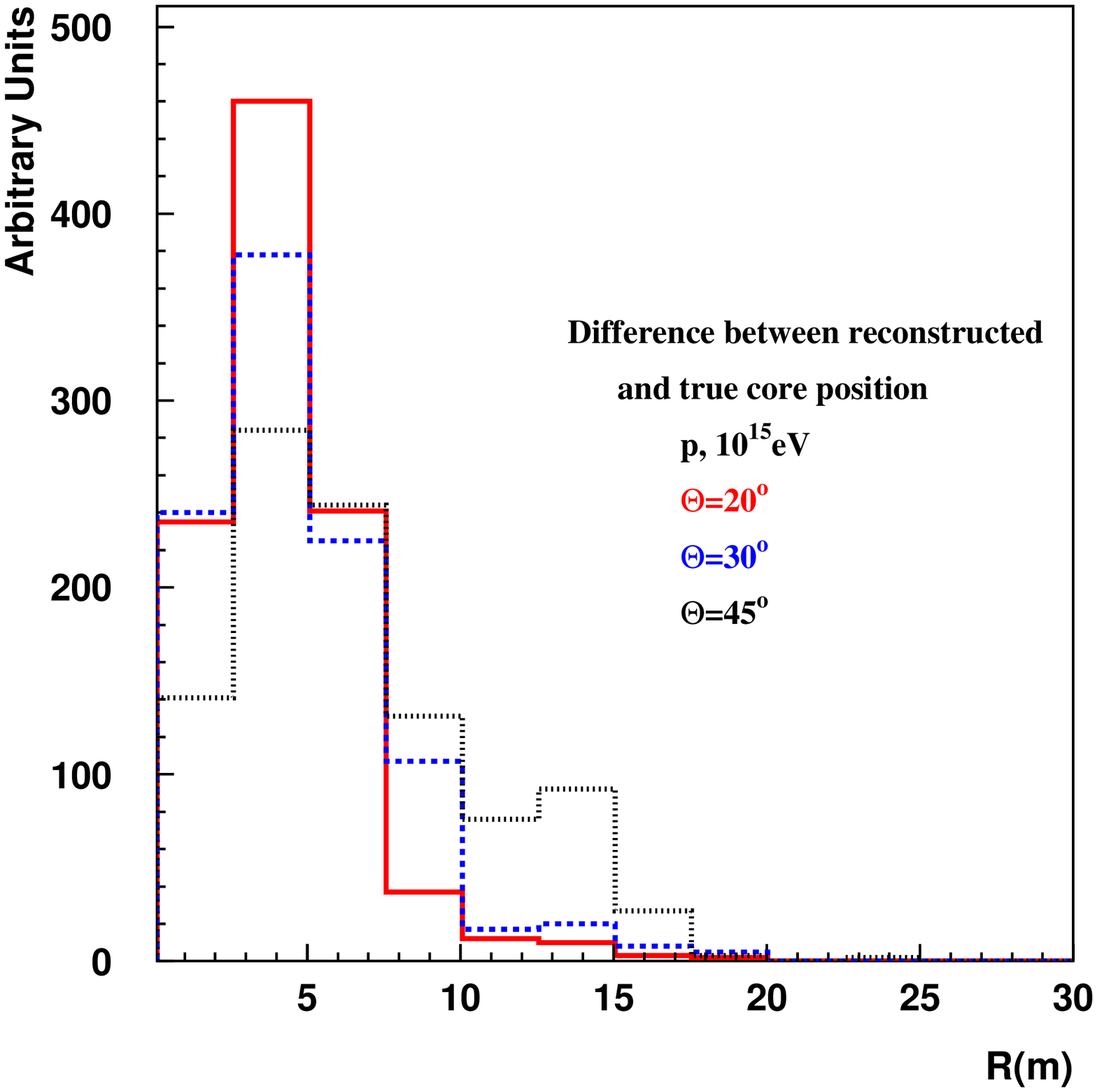}
\caption{The reconstructed shower cores for 250 H incident
showers placed in the same position (0,0) for 30$^{0}$ (left) and the
quality of the reconstruction for 750 H incident showers
uniform distributed on the array’s surface for 20, 30 and 45$^{0}$
zenithal incidences (right), of the real configuration of the
mini-array} \label{reconstruction}
\end{figure}

The simulations studies of the possible configurations of the mini-array have
been performed with CORSIKA based on QGSJET model for a mini array
with detector stations placed at different distances from WILLI detector and
with different distances between the modules. The quality of the shower core
reconstruction was checked for all the cases. Fig.\ref{reconstruction} shows the reconstructed
shower cores for 250 H incident showers placed in one position for 30$^{0}$ incident
angles (left) and the quality of reconstruction for 750 H incident showers uniform
distributed on the array’s surface for 20, 30 and 45$^{0}$ zenithal incidences
(right) \cite{19}.

\begin{figure*}[t]
\centering
\includegraphics[width=50mm]{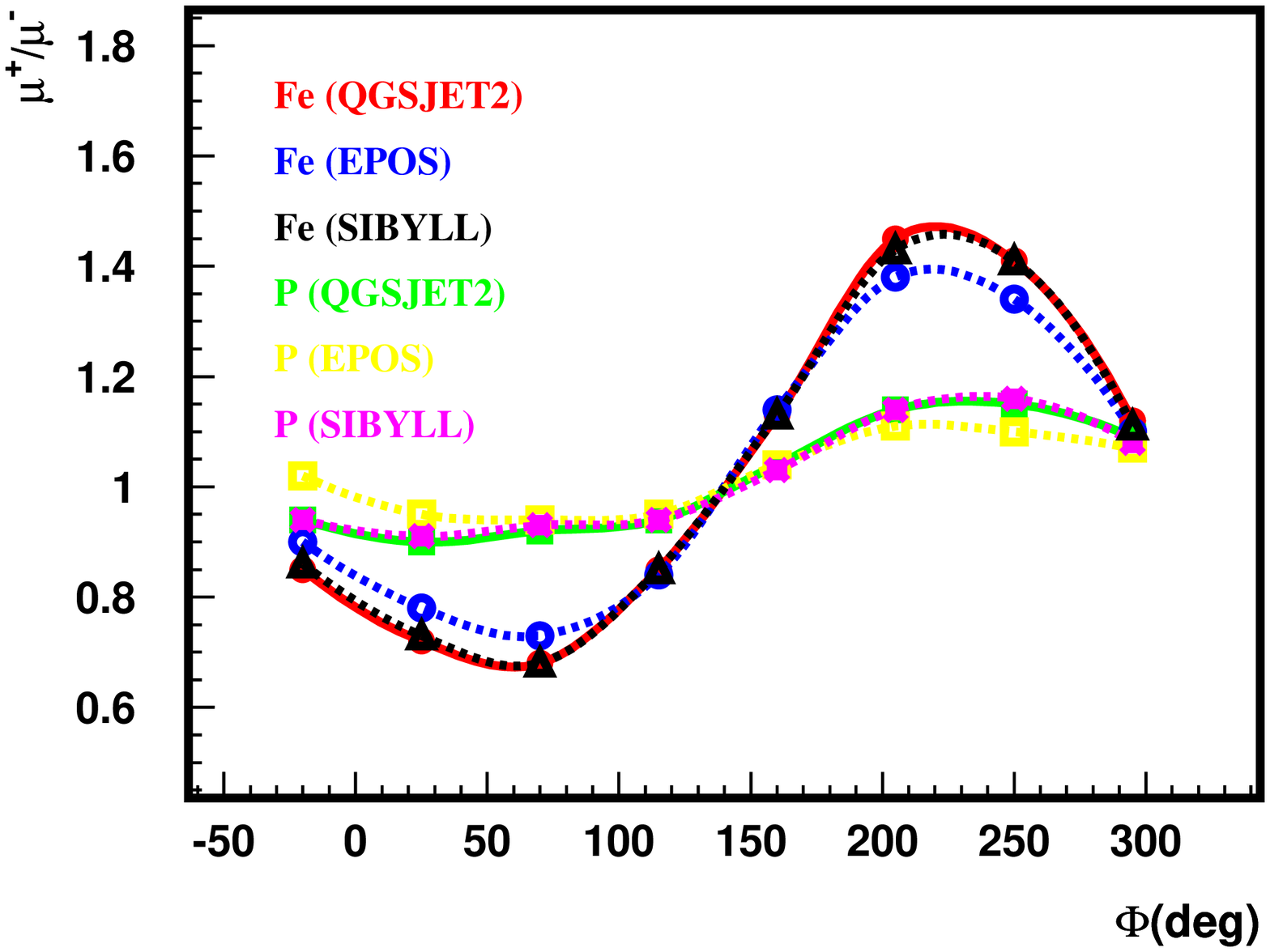}
\includegraphics[width=50mm]{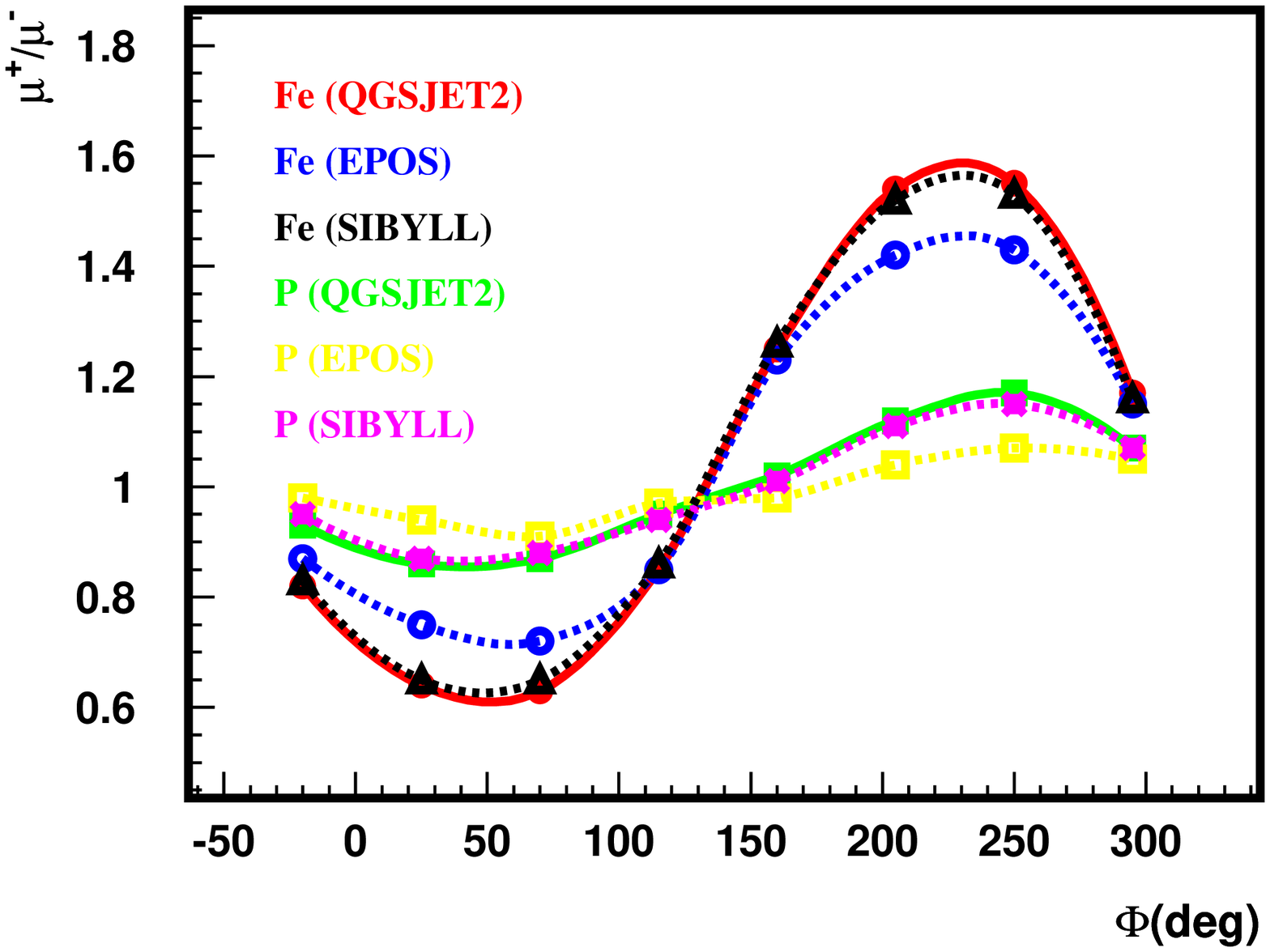}
\includegraphics[width=50mm]{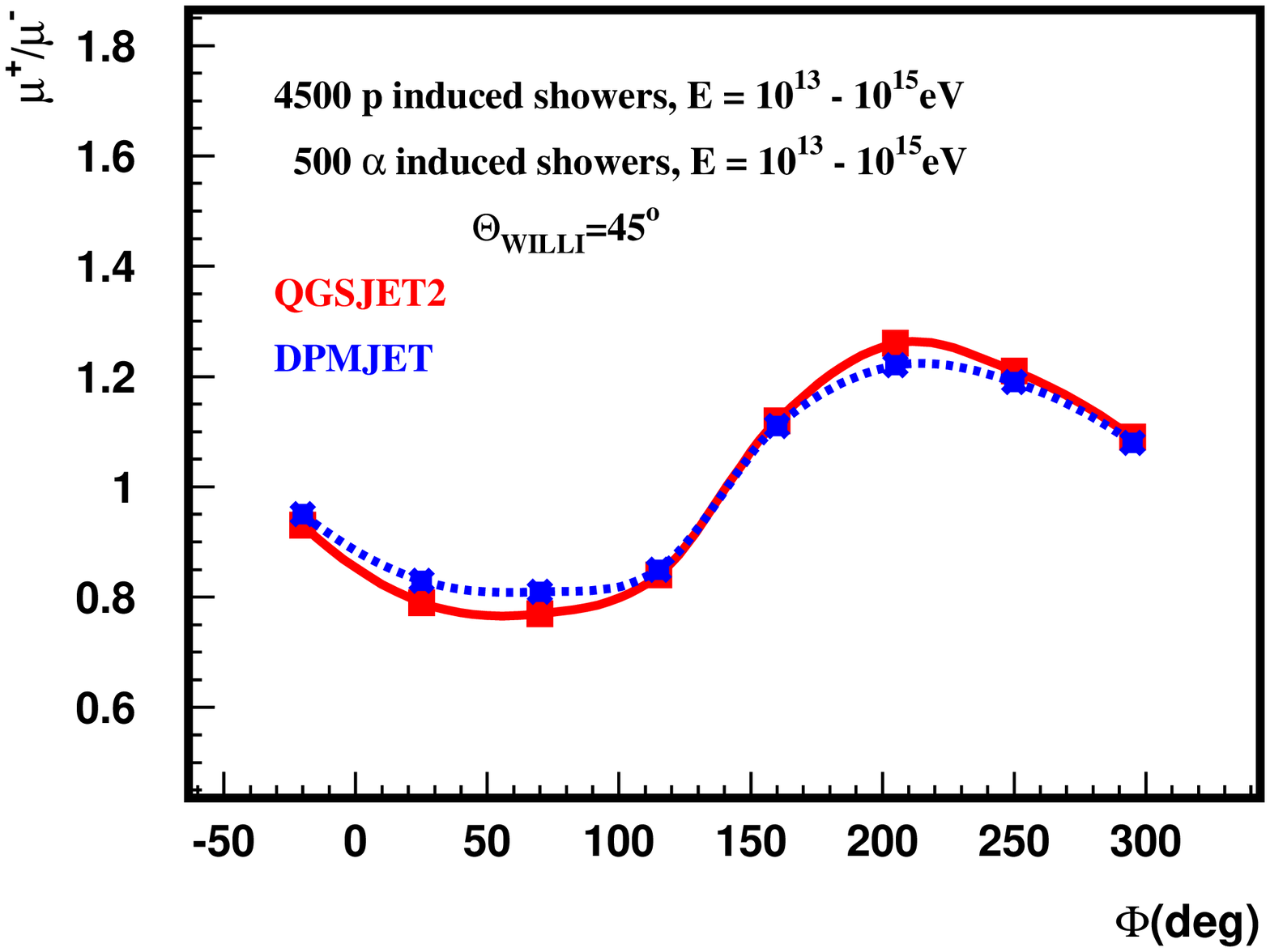}
\caption{The dependence of the charge ratio on the azimuth position of WILLI, at 15-20
m (left) and 45-50 m (center) from the shower axis, for proton and iron induced showers,
coming from South at 30$^{0}$ zenithal inclination, using 3 different hadronic interaction
models (QGSJET2, EPOS and SYBILL); The variation with the azimuthal position of
WILLI (inclined with 45$^{0}$) of the EAS muon charge ratio for 5000 incident showers with
10$^{13}$ $<$ Eprimary $<$ 10$^{15}$eV at 45-55 m distance from the shower core simulated with 2
hadronic interaction models: QGSJET2 - full line and DPMJET - dotted line. (right)} \label{azimuth}
\end{figure*}

By simulating different incident H and Fe showers, with different energies and
with different angles of incidence, the influence of the hadronic interaction
model on the EAS muon charge ratio has been investigated. Three different
interaction models, QGSJET2, EPOS and SYBILL, have been used. Fig.\ref{azimuth}
(left and center) compares the azimuth variation of muon charge ratio for the
three models

Using QGSJET2 and DPMJET interaction models, the simulations have been
performed for the real conditions of the experiment, see Fig.\ref{azimuth} (right), observing
a pronounced asymmetry with increasing angle-of-incidence.

\section{Conclusions}

i). The simulations of the muon flux for low energy ($<$ 1 GeV), using two interaction models (DPMJET and QGSJET2) have been compared with results of the measurements performed with WILLI detector, noticing good agreement for both, with very small differences.

ii). The rotatable system WILLI allows measurements of the charge ratio of muons with different angles-of incidence, being possible to determine the East-West effect of the Earth's magnetic field. The measurements with rotatable WILLI, inclined at 35$^{0}$, show a pronounced East-West effect, in good agreement with simulations data using CORSIKA code with DPMJET model, and with the East-West effect found in neutrino measurements \cite{20}.

iii). The simulations of H and Fe showers incident on WILLI-EAS detector, with different energies and with different angles of incidence, have been compared for three different interaction models, QGSJET2, EPOS and SYBILL. The results of simulations show some difference between the three models and also between H and Fe showers.

iv). The simulations of EAS muon charge ratio performed for the real condition of the experiment WILLI-EAS, using two interaction models, DPMJET and QGSJET2, indicate small differences between them, which become more pronounced for higher inclination of WILLI.

\bigskip 
\begin{acknowledgments}
The present work has been possible due to the support of the Romanian Authority for Scientific Research CNCSIS-UEFISCSU 567 by the projects: grant PNII-IDEI no.461/2009, code 1442/2008 and PNII-PARTENERIATE 82-104/2008 and project PN 09 37 01 05, and due to the strong contribution from KASCADE-Grande group from Karlsruhe Institute of Technology - KIT - Campus North, Institut fur Kernphysik, Karlsruhe, Germany.
\end{acknowledgments}

\bigskip 

\end{document}